\documentclass[conference]{IEEEtran}
\IEEEoverridecommandlockouts
\usepackage{amssymb}
\usepackage{amsmath}

\usepackage{booktabs}

\usepackage{ifpdf}

\usepackage{graphicx}

\usepackage[usenames,dvipsnames]{color}

\usepackage{mathtools}

\usepackage{epstopdf}

\usepackage{listings}

\usepackage{scalerel}

\usepackage{tikz}
\usetikzlibrary{shapes,arrows,automata,positioning,cd}
\tikzset{
  cfgedge/.style   = {black, ->, >=stealth},
  forward/.style = { blue, ->, >=angle 45},
  backward/.style = { red, densely dashed, ->, >=latex' },
  backwardleft/.style = { red, densely dashed, <-, >=latex' },
}
\usepackage{xcolor}
\usepackage[framemethod=tikz]{mdframed}

\usetikzlibrary{fit,calc}

\colorlet{pink}{red!40}
\colorlet{cyan}{cyan!60}

\usepackage{prettyref}
\newcommand{\pref}{\prettyref}
\newrefformat{thm}{Theorem~\ref{#1}}
\newrefformat{eq}{Equation~\ref{#1}}
\newrefformat{lem}{Lemma~\ref{#1}}
\newrefformat{cha}{Chapter~\ref{#1}}
\newrefformat{sec}{Section~\ref{#1}}
\newrefformat{tab}{Table~\ref{#1}}
\newrefformat{fig}{Figure~\ref{#1}}
\newrefformat{alg}{Algorithm~\ref{#1}}
\newrefformat{exa}{Example~\ref{#1}}
\newrefformat{def}{Definition~\ref{#1}}
\newrefformat{li}{Line~\ref{#1}}

\usepackage{enumitem}
\newlist{ecomponents}{enumerate}{1}
\setlist[ecomponents,1]{label={\bfseries C\arabic*},align=left}
\newlist{rqs}{enumerate}{1}
\setlist[rqs,1]{label={\bfseries RQ\arabic*},align=left}

\definecolor{light-gray}{gray}{0.9}
\definecolor{light-pink}{rgb}{0.858, 0.188, 0.478}

\newcommand{\Omit}[1]{}

\newcommand{\arxivonly}[1]{}

\newcommand{\meta}[1]{{\color{blue}\textbf{[}#1\textbf{]}}}

\newcommand{\subsubsubsection}[1]{\smallskip\noindent\textbf{\emph{#1}}\enspace}

\newcommand{\libxc}{\textsc{Libxc}}
\newcommand{\exc}{E_{\rm xc}} %
\newcommand{\ec}{E_{\rm c}} %
\newcommand{\exctilde}{\widetilde{E}_{\rm xc}} %

\newcommand{\epsxc}{{\epsilon}_{\text{xc}}} %
\newcommand{\epsx}{{\epsilon}_{\text{x}}} %
\newcommand{\epsc}{{\epsilon}_{\text{c}}} %
\newcommand{\epsxctilde}{\tilde{\epsilon}_{\text{xc}}} %
\newcommand{\epsxtilde}{\tilde{\epsilon}_{\text{x}}} %
\newcommand{\epsctilde}{\tilde{\epsilon}_{\text{c}}} %

\newcommand{\epsxpbe}{\epsxtilde^{\,\text{PBE}}}

\newcommand{\epsxunif}{\epsx^{\,\text{unif}}}
\newcommand{\uc}{U_{\text{c}}}
\newcommand{\uxc}{U_{\text{xc}}}
\newcommand{\tc}{T_{\text{c}}}
\newcommand{\clo}{C_{\text{LO}}}

\newcommand{\SAT}{\texttt{SAT}}
\newcommand{\UNSAT}{\texttt{UNSAT}}

\newcommand{\fx}{\widetilde{F}_{\text{x}}}
\newcommand{\fc}{\widetilde{F}_{\text{c}}}
\newcommand{\fxc}{\widetilde{F}_{\text{xc}}}

\usepackage{pifont}%
\newcommand{\cmark}{\ding{51}}%
\newcommand{\xmark}{\ding{55}}%
\newcommand{\timeout}{\mathord{?}}
\newcommand{\pass}{\text{\cmark}}
\newcommand{\fail}{\text{\xmark}}
\newcommand{\dreal}{\texttt{dReal}}

\newcommand{\maple}{Maple}

\usepackage{subcaption}

\newcommand{\mydot}{\rule{0.6em}{0.6em}}
\newcommand{\lightgreenbox}{\textcolor{YellowGreen}{\mydot}}
\newcommand{\lightredbox}{\textcolor{Salmon}{\mydot}}
\newcommand{\lightgreybox}{\textcolor{Gray}{\mydot}}
\newcommand{\darkgreybox}{\textcolor{Black}{\mydot}}
\newcommand{\reddot}{\textcolor{red}{\mydot}}
\newcommand{\greendot}{\textcolor{OliveGreen}{\mydot}}

\newcommand{\xcverifier}{\textsc{XCVerifier}}
\newcommand{\xcencoder}{\textsc{XCEncoder}}
\newcommand{\verifier}{\textsc{Verifier}}
\newcommand{\pb}{\textsc{PB}}

\newcommand{\cons}{\bigodot} %

\usepackage{cite}
\usepackage{amsmath,amssymb,amsfonts}
\usepackage{algorithm}
\usepackage[noend]{algorithmic}
\usepackage{graphicx}
\usepackage{textcomp}
\usepackage{xcolor}

\def\BibTeX{{\rm B\kern-.05em{\sc i\kern-.025em b}\kern-.08em
    T\kern-.1667em\lower.7ex\hbox{E}\kern-.125emX}}
\begin{document}

\title{
Towards Verifying Exact Conditions for Implementations of Density Functional Approximations
}

\author{\IEEEauthorblockN{Sameerah Helal, Zhe Tao, Cindy Rubio-Gonz\'{a}lez, Francois Gygi, Aditya V.\ Thakur}
\IEEEauthorblockA{Department of Computer Science \\
University of California, Davis\\
Davis, U.S.A. \\
\{shelal, zhetao, crubio, fgygi, avthakur\}@ucdavis.edu}
}

\maketitle

\begin{abstract}
    Density Functional Theory (DFT) is used extensively in the computation of
    electronic properties of matter, with applications in solid state physics,
    computational chemistry, and materials science. Approximating the
    exchange-correlation (XC) functional is the key to the Kohn-Sham DFT
    approach, the basis of most DFT calculations. The choice of this density
    functional approximation (DFA) depends crucially on the particular system
    under study, which has resulted in the development of hundreds of DFAs.
    Though the exact density functional is not known, researchers have
    discovered analytical properties of this exact functional. Furthermore,
    these \emph{exact conditions} are used when designing DFAs. 
    
    This paper presents \xcverifier{}, the first approach for verifying whether
    a DFA implementation satisfies the DFT exact conditions. \xcverifier{} was
    evaluated on five DFAs from the popular \libxc{} library and seven exact
    conditions used in recent work by Pederson and Burke. \xcverifier{} was able
    to verify or find violations for a majority of the DFA-condition pairs,
    demonstrating the feasibility of using formal methods to verify DFA
    implementations. However, it timed out on all conditions of the recent SCAN
    functional, revealing directions for future work. 
    \end{abstract}
    
\section{Introduction}
\label{sec:Introduction}
\emph{Density functional theory
(DFT)}~\cite{parr1995density,engel2011density,koch2015chemist} is a widely used
approximation in the computation of electronic properties of matter. DFT
calculations have been used to predict diverse properties from atomic binding
energies and chemical reactivities to electronic conductivities and magnetic
phenomena~\cite{sholl2022density}. Consequently, DFT finds applications in a
number of scientific and engineering fields, including solid state physics,
computational chemistry, and materials science, and has been 
implemented in widely-used scientific software such as ABINIT~\cite{gonze2009abinit},
cp2k~\cite{hutter2014cp2k}, ERKALE~\cite{lehtola2012erkale}, Psi4~\cite{turney2012psi4},
Octopus~\cite{castro2006octopus}, Qbox~\cite{Qbox2008}, and Quantum
ESPRESSO~\cite{giannozzi2009quantum}.

Originating in the foundational
work of Hohenberg and Kohn in 1964~\cite{KohnHohenberg1964}, DFT has provided a
practical way to reduce the complexity of conventional electronic structure
methods by identifying the electron density function $n(r)$---a real function in
three-dimensional space---as the fundamental quantity from which all other
properties of a physical system can be derived.
The Kohn-Sham (KS)~\cite{KohnSham1965} approach is currently the basis of most DFT calculations,
and states that only the exchange-correlation (XC) energy portion
of a functional needs to be approximated.
However, the exact expression for this XC energy functional $\exc[n]$, 
which describes the complex electron-electron interactions within systems, 
is not known and is incredibly difficult to approximate~\cite{ZIESCHE1998122}. 

Researchers have developed hundreds of 
approximations to the XC energy functional, which have the form: 
\begin{equation}
 \exctilde[n] =\! \int\! n(r)\, \epsxctilde\big(n(r), \nabla n(r), \nabla^2 n(r), \ldots\big)dr, 
 \label{eq:dfa}
\end{equation}
where the term $\epsxctilde$ is the \emph{density functional
approximation (DFA)}. 
The choice
of DFA in a particular application depends crucially on the known or expected
physical properties of the system under
study~\cite{becke2014perspective,mardirossian2017thirty,bursch2022best}. 

This proliferation of DFAs is reflected in the fact that the popular \libxc{}
software library~\cite{marques2012libxc}, which provides numerical implementations of DFAs, currently
includes over 500~functionals~\cite{lehtola2018recent}.
DFAs are of varying degrees of complexity~\cite{perdew2001jacob}:  (i)~local
density approximations (LDAs) depend only on the electron density~$n$,
(ii)~generalized-gradient approximations (GGAs)~\cite{perdewandwang1991ggas} depend
on the electron density $n$ and its gradient $\nabla n$, 
and (iii)~meta-GGAs have a further dependence on the Laplacian~$\nabla^2 n$ and the local kinetic 
energy density~$\tau$. 

The DFA $\epsxctilde$ in \pref{eq:dfa} is a mathematical function with a known, albeit complicated, analytical form. 
For instance, the following equation shows just the exchange~(X) part $\epsxpbe$ of the 
Perdew-Burke-Ernzerhof~(PBE) functional~\cite{perdew1996generalized}, a
commonly used GGA functional~\cite{2009whichshouldichoose}:
\begin{equation*}
    \epsxpbe(\rho, \sigma) = \frac{2.884 \rho^{1/3} \left(- 28.944 \pi^{4/3} \rho^{8/3} - 0.174 \pi^{2} \sigma\right)}{\pi^{1/3} \cdot \left(77.184 \pi^{4/3} \rho^{8/3} + 0.257 \pi^{2} \sigma\right)}
\end{equation*}
However, the correlation part of PBE is significantly more complex with over
300~operations in the \libxc{} implementation. The SCAN meta-GGA
functional~\cite{sun2015scan} is even more complex with over 1000~operations\arxivonly{ in
the exchange and correlation parts combined}, including transcendental functions 
such as $\exp$ and $\log$.

Creating a new DFA is an art mastered by only a few researchers as
of today, and the functional forms used to define the DFAs vary
considerably. DFA designs fall into two categories: empirical and non-empirical.
Empirically-designed DFAs (e.g., LYP~\cite{leeyangparr1988lyp}) are tailored for
molecular chemistry applications and perform well on molecular
benchmarks~\cite{Goerigk2017,Peverati2011}. Non-empirically designed DFAs are
constructed to satisfy some \emph{exact conditions}, which are known analytical
properties of the exact
functional~(\pref{sec:ExactConditions}). For example, the
correlation energy $E_C$ non-positivity condition states that the correlation
energy cannot be positive, i.e., $\ec[n] \leq 0$~\cite{Pederson2023}.
Furthermore, so-called norms are imposed on DFAs by requiring that they
reproduce correctly some known physical systems, e.g., a hydrogen or a helium
atom for which exact results are available. The SCAN functional is built to
satisfy as many as 17~constraints and norms~\cite{sun2015scan}.

Recently, Pederson and Burke (PB)~\cite{Pederson2023} checked whether the \libxc{}
implementations of various DFAs satisfy DFT exact conditions. In particular, they 
checked seven exact conditions by considering
their
corresponding \emph{local condition} (\pref{sec:LocalConditions}).
These local conditions were assessed for a given DFA by performing a grid search
over the inputs to the DFA and checking whether each input-output pair satisfies
the local condition. Many of the
conditions require gradient calculations, which were numerically
approximated.
The PB approach is the state of the art in the DFT community,
and was the first to perform a large scale study of the role of exact conditions
in density functional development. For instance, they found that many empirical
DFAs satisfy these exact conditions in certain regions even though they were
designed without explicit adherence to these exact conditions.

\emph{This paper addresses the problem of automatically verifying whether the
implementation of a DFA satisfies the exact conditions of the density
functional.} It is the first to apply formal-methods techniques to density
functional theory. 
The aim is to provide formal guarantees related to
the correctness of existing DFA implementations; viz., to formally verify if the
DFA implementation satisfies the exact conditions, and to determine the areas of
the input domain where it does not. 
As a step towards solving this problem, we designed and implemented
\xcverifier{}, 
a tool that verifies whether a \libxc{} functional, implemented 
in \maple{}\cite{heck1993introduction}, satisfies the given local condition 
(and, hence, the corresponding
exact condition)
using the \dreal{} solver~\cite{gao2013dreal}~(\pref{sec:Approach}). \xcverifier{} 
also computes any required derivatives symbolically, avoiding any potential 
issues arising from their numerical approximation. 
We also
implemented a domain-splitting technique to improve the performance of
\xcverifier~(\pref{alg:verify}).

We evaluated \xcverifier{} by verifying seven exact conditions (from Pederson
and Burke) for five popular DFAs: PBE~\cite{perdew1996generalized},
SCAN~\cite{sun2015scan}, LYP~\cite{leeyangparr1988lyp},
AM05~\cite{armiento2005am05}, and VWN RPA~\cite{vosko1980vwnrpa}, which cover
the different types of DFAs (LDA, GGA, and meta-GGA), as well as different
design categories (empirical and non-empirical). Some conditions do not apply to
certain DFAs, which left us with 31 DFA-condition pairs. As shown in
\pref{sec:ExperimentalEvalation-Verification}, \xcverifier{} was successfully
able to verify or find counterexamples for 13 pairs, and it is able to partially verify
an additional seven
pairs~(\pref{tab:dreal_verification_results}). These results demonstrate the
feasibility of using formal-methods techniques to verify DFA implementations.
However, \xcverifier{} times out for 11 pairs: one property for 
the PBE DFA, three properties for AM05, and \emph{all} of the properties for SCAN. 
This motivates further
research on formal methods for~DFT. 

To further validate our approach, we compared the results from the
\pb{} approach with \xcverifier~(\pref{sec:ExperimentalEvaluation-Comparison}):
both approaches find similar regions where the conditions were violated or
satisfied~(\pref{tab:pb_xcverifier_consistency}).

The contributions of the paper are as follows: 
\begin{itemize}
  \item A tool,  \xcverifier{}, for automatically verifying exact conditions for 
  density functional approximations~(\pref{sec:Approach}).
  \item An evaluation of \xcverifier{} using five DFAs and seven exact
  conditions along with a comparison with the state-of-the-art grid-search
  approach~(\pref{sec:ExperimentalEvalation}).
\end{itemize}
Our preliminary results demonstrate the feasibility of using formal methods to
prove the correctness  DFT implementations, and reveal avenues for
future work~(\pref{sec:Discussion}).

\section{Exact Conditions in DFT}
\label{sec:ExactConditions}
\label{sec:LocalConditions}

This section lists the exact conditions of the density functional $\exc$
considered in this paper. Our description closely follows that in Pederson and
Burke~\cite{Pederson2023}. Each exact condition has a corresponding \emph{local
condition} such that if the DFA $\epsxctilde$ satisfies the local condition,
then the functional $\exctilde$ satisfies the (global) exact condition.
Note that the converse is not true: violating the local condition 
does not imply that the exact condition is violated. Furthermore, 
the region where the local condition is violated for a DFA can depend
on the particular implementation of the functional.
In this paper, we focus on verifying the \emph{\libxc{} implementation 
of a DFA}.

The local conditions take the exchange (correlation) enhancement factor
$\fxc$, which is a function of $\epsxc\big(n(r)\big)$, the local value of the
exchange (correlation) energy for the~DFA. 
Here, $n(r)$ is the electron density at the point $r$ representing 
the position of an electron. For GGA functionals, the inputs $n(r)$ and $\nabla
n(r)$ are usually expressed in terms of the Wigner-Seitz radius
$r_s=(4\pi n/3)^{-1/3}$ and $s=|\nabla n|/(2(3\pi^2)^{1/3}n^{4/3})$.
Given the expressions for the DFA exchange and correlation energies
$\epsxtilde$ and $\epsctilde$ in terms of $s$ and~$r_s$, we can 
compute $\fx$ and $\fc$ of $\fxc$: 
\begin{equation}
    \fxc[n(r)] = \fx + \fc = \frac{\epsxctilde[n(r)]}{\epsxunif[n(r)]}.
\end{equation}

We now list the DFT exact conditions along with their corresponding local condition:

\noindent\textbf{(EC1) The correlation energy ($\ec$) non-positivity
    condition}~\cite{Pederson2023} is defined as
        $\ec[n] \leq 0$.
    The corresponding local condition is
    \begin{equation}
        \epsc(n(r)) \leq 0. %
        \label{eq:local-ec-no-fc}
    \end{equation}
    It can also be expressed in terms of $\fc$ as
    \begin{equation}
        \fc \geq 0.
        \label{eq:local-ec}
    \end{equation}

    \noindent\textbf{(EC2) The $\ec$ scaling inequality}~\cite{PhysRevA.32.2010,Pederson2023},
        $(\gamma-1)\ec[n_\gamma] \geq \gamma(\gamma - 1) \ec[n]$,
    has the local condition
    \begin{equation}
        \frac{\partial \fc}{\partial r_s} \geq 0. %
        \label{eq:local-ec-scaling}
    \end{equation}
    
\noindent\textbf{(EC3) The $\uc(\lambda)$ monotonicity condition}~\cite{PhysRevB.48.11638,Pederson2023} is 
        $\frac{d\uc(\lambda)}{d\lambda} \leq 0$,
    where $\uc(\lambda) = d(\lambda^2\ec[n_{1/\lambda}])/d\lambda$ represents
    the correlation energy adiabatic connection curves.
    The corresponding local condition is
    \begin{equation}
        \frac{\partial^2 \fc}{\partial r_s^2} \geq \frac{-2}{r_s} \cdot \frac{\partial \fc}{\partial r_s}.
        \label{eq:local-uc}
    \end{equation}

\noindent\textbf{(EC4) Lieb-Oxford bound}~\cite{https://doi.org/10.1002/qua.560190306,Pederson2023} is
        $\uxc \geq \clo\int d^3 r\, n(r)\, \epsxunif\big(n(r)\big)$,
    where $\uxc[n] = \exc[n] - \tc[n]$ is the potential correlation energy and
    $\clo=2.27$ is the Lieb-Oxford constant, following~\cite{Pederson2023}.
    The corresponding local condition is
    \begin{equation}
        \fxc + r_s\frac{\partial \fc}{\partial r_s} \leq \clo.
        \label{eq:local-lo}
    \end{equation} 

\noindent\textbf{(EC5) The Lieb-Oxford extension to $\exc$}~\cite{https://doi.org/10.1002/qua.560190306,Pederson2023} is
    a generalization of the Lieb-Oxford bound with $\exc$ instead of~$\uxc$~:~
        $\exc \geq \clo\int d^3 r\, n(r)\, \epsxunif\big(n(r)\big)$.
    The corresponding local condition is
    \begin{equation}
       \fxc \leq \clo
       \label{eq:local-lo-exc}
    \end{equation}

\noindent\textbf{(EC6) The $\tc$ upper bound condition}~\cite{PhysRevB.48.11638,Pederson2023} is
        $\tc[n_\gamma] \leq -\gamma \left(\frac{\partial \ec[n_\gamma]}{\partial\gamma}\right) + \ec[n_\gamma]$,
    with corresponding local condition
    \begin{equation}
        \frac{\partial \fc}{\partial r_s} \leq \frac{\fc(\infty) - \fc}{r_s}
        \label{eq:local-tc}
    \end{equation}    
    where $\fc(\infty)$ is the limit of $\fc$ as $r_s \to \infty$.

\noindent\textbf{(EC7) The \emph{conjectured} $\tc$ upper bound}~\cite{Levy1985,10.1063/1.481099,Crisostomo2023,Pederson2023} is
        $\tc[n] \leq -\ec[n]$
    with local condition
    \begin{equation}
        \frac{\partial \fc}{\partial r_s} \leq \frac{\fc}{r_s}.
        \label{eq:local-tc-conjectured}
    \end{equation}

\section{\xcverifier}
\label{sec:Approach}

This section describes the design of the \xcverifier{} tool, which verifies whether
a DFA implemented in the \libxc{} library~\cite{marques2012libxc} satisfies the
DFT local conditions in \pref{sec:ExactConditions}. 
\xcverifier{} consists of (i)~\emph{\xcencoder{}}, which encodes the
given local condition for a given \libxc{} functional into a formula~$\psi$, and
(ii)~\emph{\verifier{}}, which verifies whether this encoded formula~$\psi$ is always
true (valid) in the given input domain. Because DFAs involve non-linear
arithmetic and transcendental functions, we chose the \dreal{}
solver~\cite{gao2013dreal} as the core solver in \verifier{}. As a
consequence, the formula $\psi$ generated by \xcencoder{} is a \dreal{} formula.

\subsection{\xcencoder}

Given a \libxc{} functional, \xcencoder{} first translates the \maple{} code
for $\epsxctilde$ to Python using
the \texttt{CodeGeneration} package from \maple. 
We implemented a symbolic execution engine for (a subset of) Python
that generates the \dreal{} expression corresponding to the DFA. Though DFA implementations do
not contain loops, arrays, etc., they do contain (non-recursive) function calls
and if-then-else statements. 

\arxivonly{
To test that this translation is
correct, we uniformly sample points from the input region and assert that the
outputs of the translated \changed{\dreal{}} implementation and the original \libxc{}
implementation are close by a factor of at least $10^{-13}$. 
\changed{Of the DFAs we use
in our experiments, our translations are within $10^{-14}$
of their corresponding \libxc{} implementations for all points sampled for PBE, AM05, and VWN RPA;
and within $10^{-13}$ for LYP. Our translation of SCAN is within
$10^{-14}$ for $98\%$ of the points sampled. The remaining $2\%$ of
the points are within $10^{-1}$ of the \libxc{} implementation,
which is a dramatic difference \meta{this is a $99\%$ relative difference}. 
This may be due to the extreme complexity of the SCAN functional. In our experiments,
we observed no pattern in the points where the translations did not match.
}
} 

\xcencoder{} then constructs the \dreal{} formula $\psi$ that encodes the given
local condition for the particular functional.
Encoding the local condition corresponding to the $\ec$ non-positivity condition
is straightforward: \xcencoder{} uses the \dreal{} expression for $\epsctilde$
to directly construct the \dreal{} formula $\epsctilde \leq 0$. 

However, the local conditions corresponding to exact conditions such as $\ec$
scaling, Lieb-Oxford, $\uc$ monotonicity, and $\tc$ upper bound require computation
of one or more derivatives, In such cases, \xcencoder{} uses SymPy~\cite{Meurer2017} 
to symbolically compute the derivatives. Furthermore, the
local condition corresponding to the $\tc$ upper bound condition~(\pref{eq:local-tc}) requires computing
 $\lim_{r_s\to\infty}\fc$. 
Following~\cite{Pederson2023}, \xcencoder{} substitutes an appropriately large value 
to approximate this limit at infinity, viz.\ $\fc|_{r_s=100}$.

\subsection{\verifier}
\label{sec:Approach-VerifyingLocalConditions}

\begin{figure}
  \begin{algorithm}[H]
    \caption{\verifier(input domain $D$, formula $\psi$)}
    \label{alg:verify}
    \begin{algorithmic}[1]
      \IF {$D < t$}    
         \RETURN \label{li:toosmall}
      \ENDIF
      \STATE result, x $\gets$ \dreal($\varphi_D \wedge \lnot \psi$) \label{li:dreal}
      \IF {result $=$ \UNSAT} 
         \PRINT ``Verified condition over domain $D$''
         \RETURN \label{li:unsat}
      \ENDIF
      \IF {result $=$ \SAT}
          \IF {\texttt{valid}(x)}  \label{li:isvalid}
              \PRINT ``Found counterexample x'' \label{li:counterexample}
          \ELSE 
              \PRINT ``Verification inconclusive over domain $D$'' \label{li:inconclusive}
          \ENDIF
      \ELSE 
           \PRINT ``Verification timed out over domain $D$'' \label{li:timeout}
      \ENDIF
      \FORALL{$D'$ \textbf{in} \texttt{split}($D$)} \label{li:split}
        \STATE \verifier($D'$, $\psi$) \label{li:recurse}
      \ENDFOR
      \RETURN 
    \end{algorithmic}
    \end{algorithm}
    \vspace{-3ex}
  \end{figure}
  
\pref{alg:verify} presents the pseudo-code for \verifier{}. The \verifier{}
component of \xcverifier{} takes as input (i)~the formula $\psi$ encoding the
exact condition for the DFA, and (ii)~a domain for the inputs to the DFA. We use
the same input bounds as used in Pederson and Burke~\cite{Pederson2023}; for
example, for GGA functionals, the domain for $r_s$ is the interval $[0.0001,
5]$, and that for $s$ is $[0,5]$. Consequently, for a GGA functional, the
\verifier{} is trying to prove the validity, or satisfiability, of the following
formula:
\begin{equation}
\forall r_s, s\,.\ (r_s \in [0.0001, 5] \wedge s \in [0,5]) \implies \psi.
\label{eq:forall-formula}
\end{equation}
Proving the validity of \pref{eq:forall-formula} is equivalent to proving that 
the following formula is unsatisfiable:
\begin{equation}
  r_s \in [0.0001, 5] \wedge s \in [0,5] \wedge \lnot \psi.
  \label{eq:exists-formula}
\end{equation}

\verifier{} uses the \dreal{} solver~\cite{gao2013dreal} to check the
satisfiability of the above formula. 

\dreal{} implements a delta-complete
decision framework: given a formula $\varphi$, \dreal{} returns
\UNSAT{}---$\varphi$ is unsatisfiable---or $\delta$-\SAT{}---the
$\delta$-weakening $\varphi^{\delta}$ is satisfiable for the returned model,
where a model is a satisfying assignment to the variables in the formula.
The
$\delta$-weakening $\varphi^{\delta}$ of a formula $\varphi$ is numerical
relaxation of $\varphi$ such that (i)~a model that satisfies $\varphi$ will
always satisfy $\varphi^{\delta}$; however, the reverse is not necessarily true,
and (ii)~if $\varphi^{\delta}$ is unsatisfiable, then $\varphi$ is also
unsatisfiable. This relaxation results in \dreal{} now being decidable for
nonlinear formulas including those with transcendental functions, which are
common in DFAs. 

Though the $\delta$-satisfiability problem is decidable in principle, in
practice, the \dreal{} solver could also \emph{timeout}: it was unable to
determine whether formula $\varphi$ is \UNSAT{} or $\delta$-\SAT{} in the
given time limit. Our preliminary results showed that \dreal{} would timeout on
the formula in \pref{eq:exists-formula} for most functional/condition pairs, even when given a time limit of
24~hours. To improve the performance of \verifier{}, we implement a
domain-splitting technique that partitions the input domain and uses \dreal{} to
solve the formula $\lnot \psi$ on each subdomain separately. This simple
strategy greatly improves the performance of \verifier. Furthermore,
one of the goals of \xcverifier{} is to determine the input regions where the
DFA implementation violates the local condition. Thus, \verifier{} also performs
the domain-split when \dreal{} returns a valid model: an input that indeed
violates the given condition. This allows \verifier{} to isolate the subregions
where the DFA implementation violates the local condition.

\pref{alg:verify} presents the pseudo-code for \verifier{}, and 
\pref{fig:PBE_T_C_upper_bound_conjectured_solver} shows a graphical representation of
the output of \verifier{} when verifying the conjectured $\tc$ upper bound condition for the PBE functional.

The \dreal{} solver is called on \pref{li:dreal} to find a satisfying assignment
to the formula $\varphi_D \wedge \lnot\psi$, where $\varphi_D$ is the formula
encoding the domain constraints on the inputs to the DFA and $\psi$ encodes the
local condition for the functional.
If the result is \UNSAT{}, then \verifier{} returns after recording that the
condition was verified over domain $D$ (\pref{li:unsat}). This is indicated
using \lightgreenbox{} in \pref{fig:PBE_T_C_upper_bound_conjectured_solver}.
If the result is $\delta$-\SAT{}, \dreal{} will also return a model $x$ for the
formula. In \pref{li:isvalid}, \texttt{valid(x)} checks if this model is a valid counterexample
by plugging the values back into $\psi$, which encodes the local condition. If
the condition is indeed violated, then \verifier{} records it as a
counterexample (\pref{li:counterexample}), indicated using \lightredbox{}.
Occasionally, \dreal{} may return \SAT{} with a model that does not actually
violate $\psi$. This happens due to the $\delta$-satisfiability procedure of
\dreal{}: the model must satisfy the weakened formula $(\lnot\psi)^{\delta}$,
but not necessarily the original formula $\lnot\psi$. In such cases, \verifier{}
records the result as inconclusive (\pref{li:inconclusive}), indicated using
\darkgreybox{} in  
\pref{fig:PBE_T_C_upper_bound_conjectured_solver}.
We use a two hour time limit for the \dreal{} solver; if \dreal{} is unable to
determine (un)satisfiability in this limit, \verifier{} interrupts it and
records that the verification timed out for domain~$D$ (\pref{li:timeout}),
indicated using \lightgreybox{} in \pref{fig:PBE_T_C_upper_bound_conjectured_solver}.
Increasing the timeout in our experiments did not enable \dreal{} to solve
more formulas.

\pref{li:split} is executed if the result is \SAT{} or when \dreal{} times out.
\verifier{} calls \texttt{split}($D$), which partitions each input dimension of
$D$ into two equal parts. \verifier{} is recursively called on each
subdomain $D'$ in \pref{li:recurse}. We set a lower limit on the size of input
domain as the base case for the recursion. On \pref{li:toosmall}, \verifier{}
returns if the given input domain is too small as determined by the threshold~$t$; 
we used $t=0.05$ in our experiments. %

\begin{table*}[t]
    \centering    
    \caption{Verifying local conditions for DFT exact conditions for DFAs using \xcverifier{}.\\
    $\pass$: \xcverifier{} verified that the DFA satisfies the condition on the entire input domain; \\
    $\pass^*$: \xcverifier{} verified that the DFA satisfies the condition on
    part of the input domain with the rest timing out/inconclusive;\\
    $\timeout$: \xcverifier{} reported timeout/inconclusive for all of the input domain;\\
    $-$: the condition does not apply to the DFA;\\
    $\fail$: \xcverifier{} found a counterexample showing that the DFA does not satisfy the local condition.
    }
    \begin{tabular}{lccccc}
        \toprule
        Local condition & \textbf{PBE} & \textbf{LYP} & \textbf{AM05} & \textbf{SCAN} & \textbf{VWN RPA}\\
        \midrule
        \textbf{$\ec$ non-positivity} (\pref{eq:local-ec})                 & $\pass^*$ & $\fail$ & $\pass$ & $\timeout$ & $\pass$ \\
        \textbf{$\ec$ scaling inequality} (\pref{eq:local-ec-scaling})     & $\pass^*$ & $\fail$ & $\pass^*$ & $\timeout$ & $\pass$  \\
        \textbf{$\uc$ monotonicity} (\pref{eq:local-uc})                   & $\timeout$ & $\fail$ & $\timeout$ & $\timeout$ & $\pass$ \\
        \textbf{$\tc$ upper bound} (\pref{eq:local-tc})                    & $\pass^*$ & $\fail$ & $\pass$ & $\timeout$ & $\pass$ \\
        \textbf{Conjectured $\tc$ upper bound} (\pref{eq:local-tc-conjectured})  & $\fail$ & $\fail$ & $\pass^*$ & $\timeout$ & $\pass^*$ \\
        \textbf{LO bound} (\pref{eq:local-lo})                         & $\pass^*$ & $-$ & $\timeout$ & $\timeout$ & $-$  \\
        \textbf{LO extension to $E_{xc}$} (\pref{eq:local-lo-exc})             & $\pass$ & $-$ & $\timeout$ & $\timeout$ & $-$ \\
        \bottomrule
    \end{tabular}
    \label{tab:dreal_verification_results}
\end{table*}
\section{Experimental Results}
\label{sec:ExperimentalEvalation}

This section evaluates the performance of \xcverifier{} for verifying local
conditions for DFA implementations in \libxc{}, and compares the results to the prior \pb{}
approach~\cite{Pederson2023}. 

Our experiments were designed to answer the following research questions:

\begin{rqs}
\item Is \xcverifier{} able to verify or find counterexamples for local conditions of DFA implementations
(\pref{sec:ExperimentalEvalation-Verification})?
\item How does \xcverifier{} compare to the \pb{}~approach
(\pref{sec:ExperimentalEvaluation-Comparison})?
\end{rqs}

\subsection{Experimental Setup}
We use the following five DFAs in our experiments: PBE~\cite{perdew1996generalized}, a popular
non-empirical GGA DFA; SCAN~\cite{sun2015scan}, a fully constrained
non-empirical meta-GGA DFA satisfying all known properties of DFAs;
LYP~\cite{leeyangparr1988lyp}, and empirical DFA that is a key component of
several commonly-used DFAs; AM05~\cite{armiento2005am05}, which shows efficient
and superior performance on solids~\cite{mattsonandarmiento2008am05}; and VWN
RPA~\cite{vosko1980vwnrpa}, an LDA functional.

For each of these DFAs, we consider each of the applicable conditions
from~\pref{sec:ExactConditions}. Note that the Lieb-Oxford conditions only apply
to functionals with both an exchange and correlation component available (e.g.,
PBE, AM05 SCAN).

The \pb{} approach is the state of the art for assessing condition satisfaction
in DFT. For a given DFA and condition, the \pb{} approach draws $10^5$ uniform
samples each for $r_s$ and $s$, which are then meshed into a grid. \pb{} then calls
the \libxc{}
implementation of the DFA for each of the points in the grid. This
grid is used to numerically compute the limits and gradients necessary for the
conditions using the NumPy package in Python~\cite{harris2020array}. Then the
condition is checked at each point in the grid. The condition is assumed to be
satisfied for the DFA if all the points in the grid pass the condition.

\subsection{Verifying Local Conditions using \xcverifier{}}

\label{sec:ExperimentalEvalation-Verification}

\begin{figure*}[t]
    \begin{subfigure}[b]{0.3\textwidth}
        \centering
        \includegraphics[width=\textwidth]{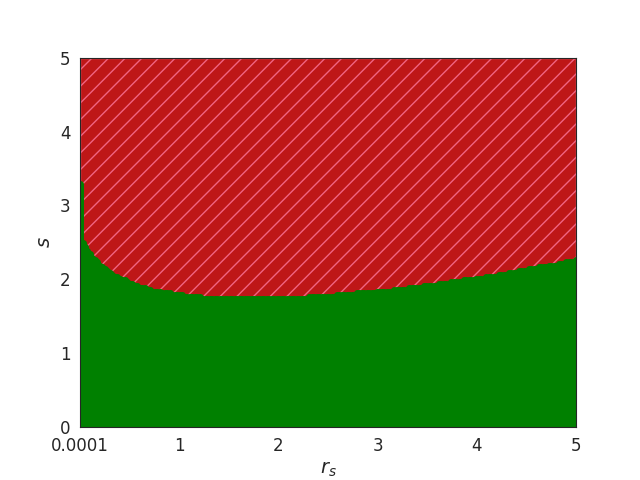}
        \caption{$\ec$ non-positivity w.\ \pb{}}
        \label{fig:LYP_E_C_negativity_grid_search}
    \end{subfigure}
    \hfill
    \begin{subfigure}[b]{0.3\textwidth}
        \centering
        \includegraphics[width=\textwidth]{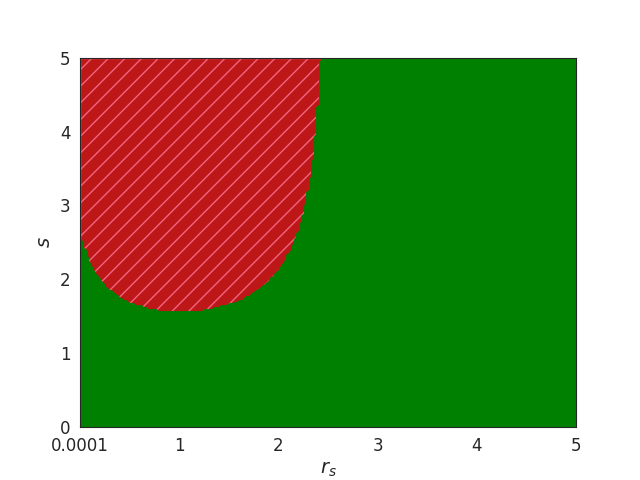}
        \caption{$\ec$ scaling inequality w.\ \pb{}}
        \label{fig:LYP_E_C_scaling_inequality_grid_search}
    \end{subfigure}
    \hfill
    \begin{subfigure}[b]{0.3\textwidth}
        \centering
        \includegraphics[width=\textwidth]{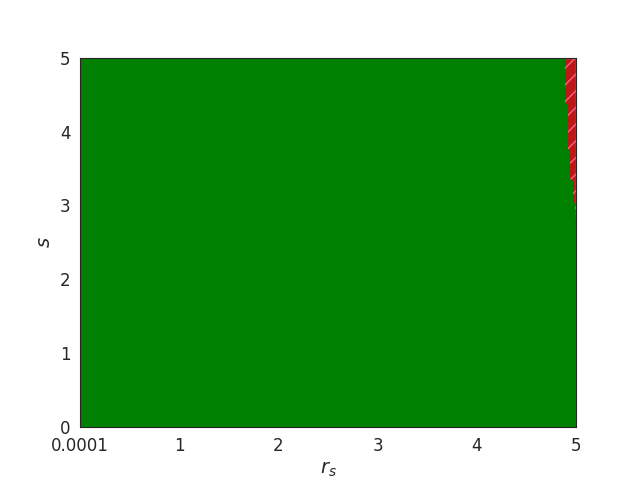}
        \caption{$\tc$ upper bound w. \pb{}}
        \label{fig:LYP_T_C_upper_bound_grid_search}
    \end{subfigure}

    \begin{subfigure}[b]{0.3\textwidth}
        \centering
        \includegraphics[width=\textwidth]{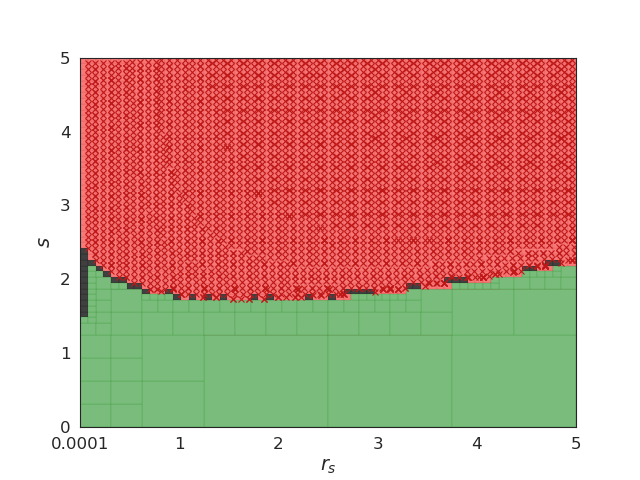}
        \caption{$\ec$ non-positivity w.\ \xcverifier{}}
        \label{fig:LYP_E_C_negativity_solver}
    \end{subfigure}
    \hfill
    \begin{subfigure}[b]{0.3\textwidth}
        \centering
        \includegraphics[width=\textwidth]{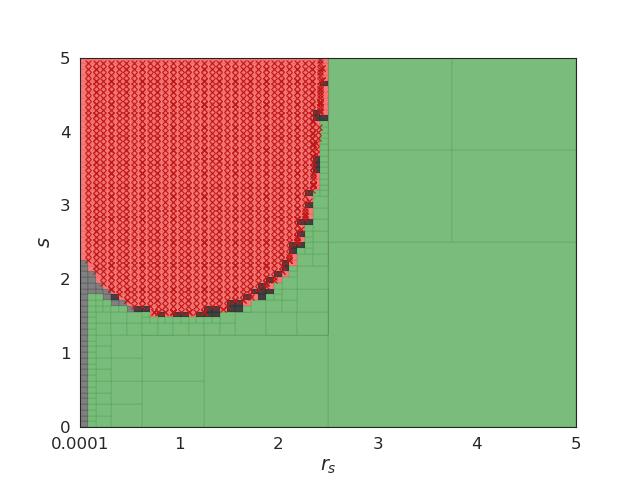}
        \caption{$\ec$ scaling inequality w.\ \xcverifier{}}
        \label{fig:LYP_E_C_scaling_inequality_solver}
    \end{subfigure}
    \hfill
    \begin{subfigure}[b]{0.3\textwidth}
        \centering
        \includegraphics[width=\textwidth]{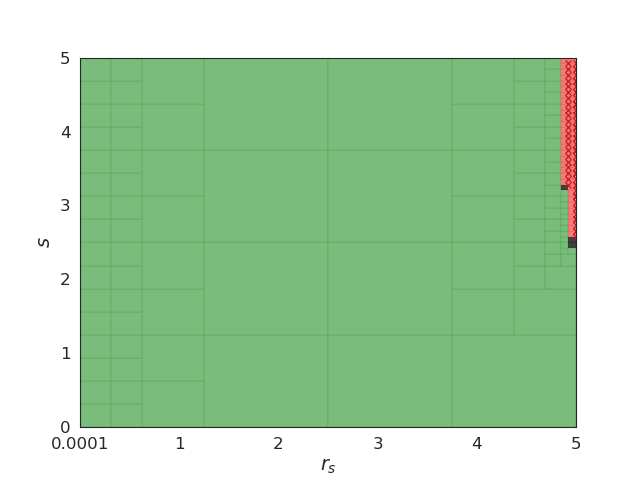}
        \caption{$\tc$ upper bound w.\ \xcverifier{}}
        \label{fig:LYP_T_C_upper_bound_solver}
    \end{subfigure}
    \caption{Regions where the LYP functional satisfies or violates conditions according to \pb{} (top) and \xcverifier{} (bottom).\\
    For \pb{}: \reddot{} (region hatched) is a counterexample to the condition, \greendot{} is a point that satisfies the condition. \\
    For \xcverifier{}: \lightredbox{} is a region that contains a counterexample marked with \textcolor{Maroon}{\texttimes}, \lightgreenbox{} is a region that is verified to satisfy the condition, 
    \lightgreybox{} indicates a timeout, and \darkgreybox{} indicates an inconclusive result.}
    \label{fig:dreal_verification_results_lyp}
\end{figure*}

\pref{tab:dreal_verification_results} summarizes the result 
of using \xcverifier{} to
verify the local condition corresponding to each DFA exact condition
in~\pref{sec:ExactConditions} for the five DFAs.
The results described below are summarized in
Table~\ref{tab:dreal_verification_results}. Visualizations for LYP and PBE are
shown in~Figures~\ref{fig:dreal_verification_results_pbe}
and~\ref{fig:dreal_verification_results_lyp}, respectively.

\subsubsubsection{PBE} 
For the $\ec$ non-positivity condition and the Lieb-Oxford bound, \xcverifier{}
is able to verify the entire input domain except for some timed-out and inconclusive
regions along the $s$-axis~(\pref{fig:PBE_E_C_negativity_solver}). The verified region is
$r_s>0.9375$ for the $\ec$ non-positivity condition and $r_s>0.0781$ for the Lieb-Oxford bound.
The Lieb-Oxford \emph{extension} to $E_{xc}$ is verified for the entire input domain~(\pref{fig:PBE_Lieb_Oxford_extension_solver}).
For the $\ec$ scaling inequality, \xcverifier{} verifies most of the bottom third of the input domain and 
times out for the rest. 
There are also a few small inconclusive regions for high $s$.
For the $\tc$ upper bound condition, \xcverifier{} verifies the bottom two-thirds of the input domain
and times out for the rest.

For the conjectured $\tc$ upper bound, \xcverifier{} finds a large
counterexample region covering the upper left diagonal of the input domain
(\pref{fig:PBE_T_C_upper_bound_conjectured_solver}). There is an
inconclusive region along the border of the counterexample region.
\xcverifier{} verifies or times out for the rest of the input domain.

Finally, \xcverifier{} times out everywhere for the $\uc$ monotonicity
condition. 

\subsubsubsection{LYP} For the $\ec$
non-positivity condition, \xcverifier{} finds counterexamples at
$s>1.6563$~(\pref{fig:LYP_E_C_negativity_solver}) and the remainder of the input domain is verified to satisfy the condition.
For the $\ec$ scaling
inequality, the counterexamples are at $r_s<2.5$ and
$s>1.4844$~(\pref{fig:LYP_E_C_scaling_inequality_solver}).
For the $\tc$ upper
bound condition~(\pref{fig:LYP_T_C_upper_bound_solver}), the counterexamples are in a small region at $r_s>4.8437$ and
$s>2.4219$, and for the conjectured $\tc$ upper bound condition, the region is $r_s>0.625$ and $s>1.3281$.
For the $\uc$ monotonicity condition, \xcverifier{} finds counterexamples
at $s>1.4844$ and $r_s < 1.4062$. The rest of the region is partially verified or timed-out.
There are some small inconclusive or timed-out regions on the borders of 
the counterexample regions for each of the conditions.

LYP is the only DFA where \xcverifier{} finds counterexamples to all applicable properties.

\subsubsubsection{AM05} \xcverifier{} verifies that AM05
satisfies the $\ec$ non-positivity and the $\tc$ upper bound conditions in the entire input domain.
It also verifies most of the domain for the
$\ec$ scaling inequality and the conjectured $\tc$ upper bound conditions, but times out along the $s$-axis 
at $r_s<0.0781$ for the $\ec$ scaling inequality and $r_s<0.1562$ for conjectured $\tc$ upper bound.
For the $\uc$ monotonicity condition, Lieb-Oxford bound, and Lieb-Oxford extension to $E_{xc}$,
\xcverifier{} times out
for the entire input domain.

\subsubsubsection{SCAN} For the SCAN functional, \xcverifier{} times out for
\emph{all} of the conditions. %

\subsubsubsection{VWN RPA} \xcverifier{} verifies that VWN RPA
satisfies the $\ec$ non-positivity, $\ec$ scaling inequality, and
$\tc$ upper bound conditions 
for the entire input domain. It also verifies the $\uc$ monotonicity condition where it timed out for several of the other functionals.
For the conjectured $\tc$ upper bound, it verifies the entire region except along the $s$-axis at $r_s<0.0781$,
where it returns inconclusive.

\subsubsubsection{Summary for \textbf{RQ1}:}
In our evaluation of the seven exact conditions for the 
five DFAs,
\xcverifier{} was able to verify or find counterexamples for 13 condition-DFA pairs
and partially verify seven, as shown in~\pref{tab:dreal_verification_results}. 
This demonstrates the feasibility of using formal-methods techniques like 
\xcverifier{} in verifying DFA implementations.
However, \xcverifier{} also timed out for 11 of the 31 applicable pairs,
particularly for the SCAN functional, 
which was designed to satisfy all known properties of DFAs.
Thus, there is room for improvement in formal-methods
techniques for DFT.

\subsection{Comparing \xcverifier{} to \pb{}}
\label{sec:ExperimentalEvaluation-Comparison} 

\begin{table*}[t]
    \centering    
    \caption{Comparison between results for \xcverifier{} and \pb{} approach. \\ %
    $\cons$: results of \pb{} are consistent with \xcverifier{};
    $\cons^*$: results of \pb{} are not inconsistent with \xcverifier{};\\
    $\, -$: condition does not apply to DFA; $\timeout$: \xcverifier{} times out. }
    \begin{tabular}{llllll}
        \toprule
        Local condition & \textbf{PBE} & \textbf{LYP} & \textbf{AM05} & \textbf{SCAN} & \textbf{VWN RPA}\\
        \midrule
        \textbf{$\ec$ non-positivity} (\pref{eq:local-ec})                      &$\cons^*$  &$\cons$    &$\cons^*$  &$\ \timeout$  &$\cons^*$  \\
        \textbf{$\ec$ scaling inequality} (\pref{eq:local-ec-scaling})          &$\cons^*$  &$\cons$    &$\cons^*$  &$\ \timeout$  &$\cons^*$  \\
        \textbf{$\uc$ monotonicity} (\pref{eq:local-uc})                        &$\ \timeout$  &$\cons$  &$\ \timeout$  &$\ \timeout$  &$\cons^*$  \\
        \textbf{$\tc$ upper bound} (\pref{eq:local-tc})                         &$\cons^*$  &$\cons$  &$\cons^*$  &$\ \timeout$  &$\cons^*$  \\
        \textbf{Conjectured $\tc$ upper bound} (\pref{eq:local-tc-conjectured}) &$\cons$  &$\cons$    &$\cons^*$  &$\ \timeout$  &$\cons^*$  \\
        \textbf{LO bound} (\pref{eq:local-lo})                                  &$\cons^*$  &$\, -$        &$\ \timeout$  &$\ \timeout$  &$\, -$        \\
        \textbf{LO extension to $E_{xc}$} (\pref{eq:local-lo-exc})              &$\cons^*$  &$\, -$        &$\ \timeout$  &$\ \timeout$  &$\, -$        \\
        \bottomrule
    \end{tabular}
    \label{tab:pb_xcverifier_consistency}
\end{table*}

\begin{figure*}
    \centering
    \begin{subfigure}[b]{0.3\textwidth}
        \centering
        \includegraphics[width=\textwidth]{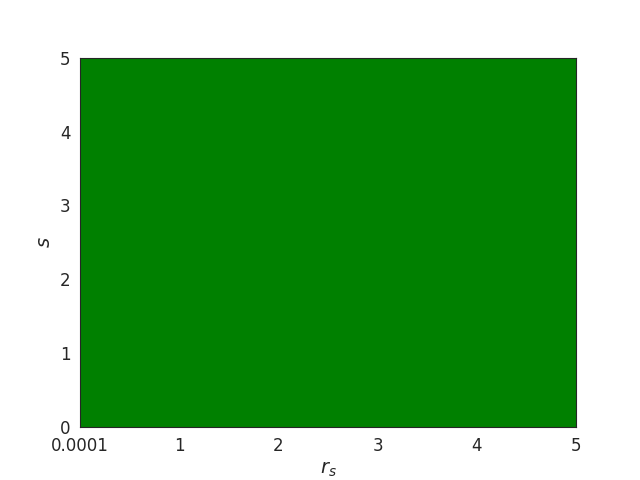}
        \caption{$\ec$ non-positivity w.\ \pb{}}
        \label{fig:PBE_E_C_negativity_grid_search}
    \end{subfigure}
    \hfill
    \begin{subfigure}[b]{0.3\textwidth}
        \centering
        \includegraphics[width=\textwidth]{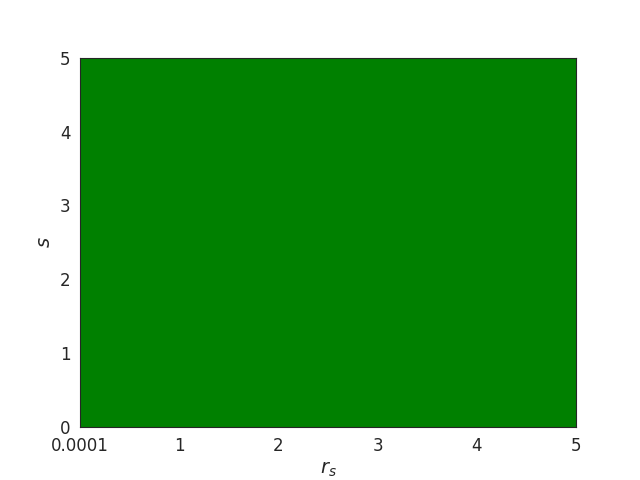}
        \caption{Lieb-Oxford ext. w.\ \pb{}}
        \label{fig:PBE_Lieb_Oxford_extension_grid_search}
    \end{subfigure}
    \hfill
    \begin{subfigure}[b]{0.3\textwidth}
        \centering
        \includegraphics[width=\textwidth]{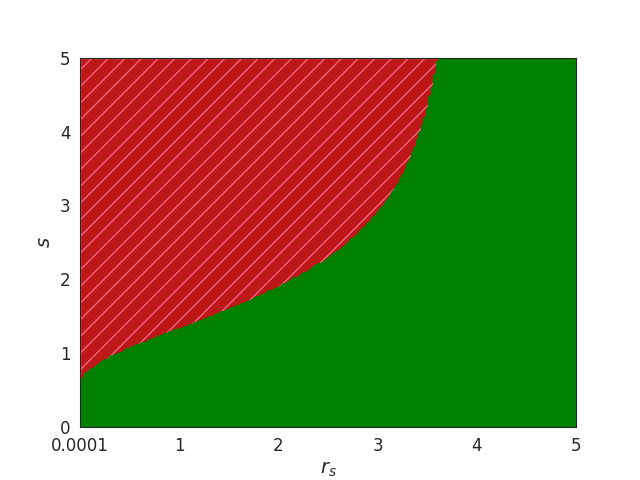}
        \caption{Conj. $\tc$ upper bound w.\ \pb{}}
        \label{fig:PBE_T_C_upper_bound_conjectured_grid_search}
    \end{subfigure}

    \centering
    \begin{subfigure}[b]{0.3\textwidth}
        \centering
        \includegraphics[width=\textwidth]{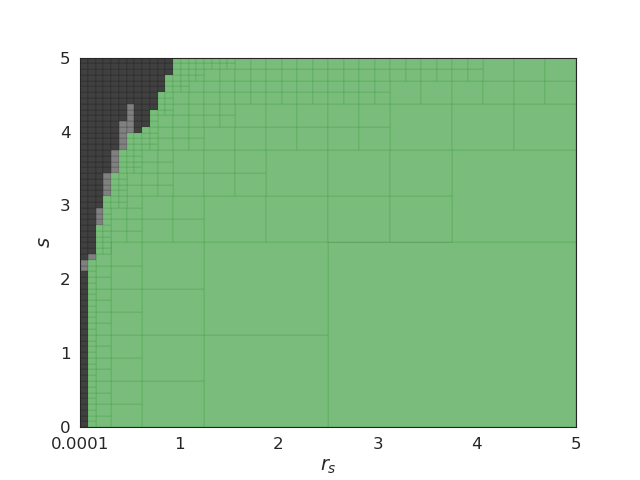}
        \caption{$\ec$ non-positivity w.\ \xcverifier{}}
        \label{fig:PBE_E_C_negativity_solver}
    \end{subfigure}
    \hfill
    \begin{subfigure}[b]{0.3\textwidth}
        \centering
        \includegraphics[width=\textwidth]{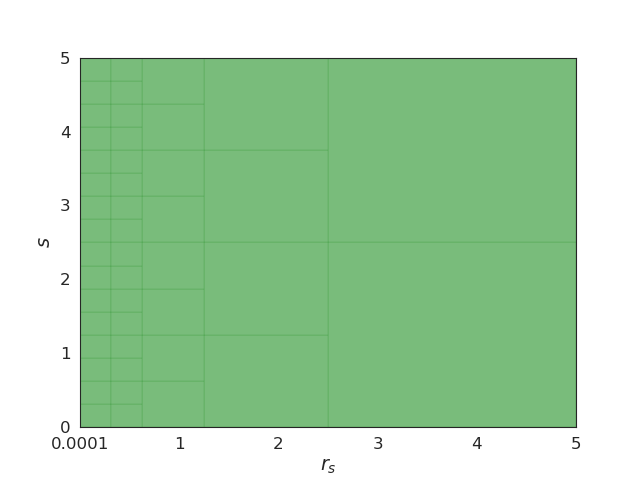}
        \caption{Lieb-Oxford ext. w.\ \xcverifier{}}
        \label{fig:PBE_Lieb_Oxford_extension_solver}
    \end{subfigure}
    \hfill
    \begin{subfigure}[b]{0.3\textwidth}
        \centering
        \includegraphics[width=\textwidth]{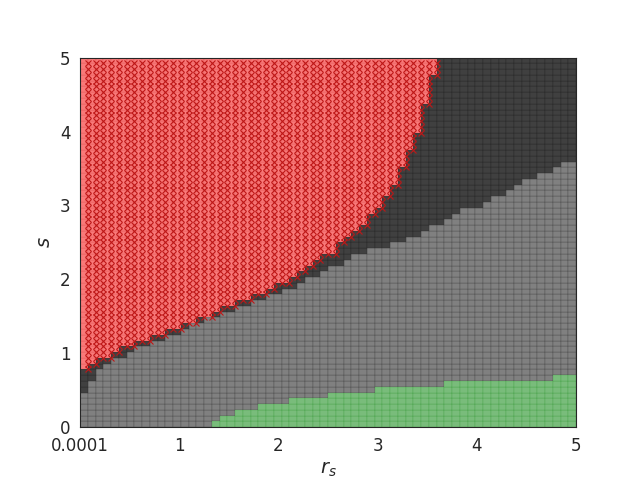}
        \caption{Conj. $\tc$ upper bound w.\ \xcverifier{}}
        \label{fig:PBE_T_C_upper_bound_conjectured_solver}
    \end{subfigure}
    \caption{Regions where the PBE functional satisfies or violates conditions according to \pb{} (top) and \xcverifier{} (bottom).\\
    For \pb{}: \reddot{} (region hatched) is a counterexample to the condition, \greendot{} is a point that satisfies the condition. \\
    For \xcverifier{}: \lightredbox{} is a region that contains a counterexample marked with \textcolor{Maroon}{\texttimes}, \lightgreenbox{} is a region that is verified to satisfy the condition,
    \lightgreybox{} indicates a timeout, and \darkgreybox{} indicates an inconclusive result.}
    \label{fig:dreal_verification_results_pbe}
\end{figure*}

This section compares the results for \xcverifier{} to the \pb{} approach.
~\pref{tab:pb_xcverifier_consistency} shows the consistency between the results
of \pb{} and \xcverifier{}. 

Out of the five functionals verified, \xcverifier{} returns counterexample regions consistent, $\cons$, with \pb{} for 
the LYP DFA
for all applicable properties. It also finds that the counterexample regions 
are consistent for the conjectured $\tc$ upper bound for PBE.

Several of the results are \emph{not inconsistent}, marked with $\cons^*$, meaning that 
neither method finds counterexamples for the condition. This is the case for when 
\pb{} finds no counterexamples,
and \xcverifier{} either verifies the entire region or partially verifies and
partially times out (e.g., for PBE in~\pref{fig:PBE_E_C_negativity_solver}).

Due to the timeouts of \xcverifier{} for all of the properties for SCAN (\pref{tab:dreal_verification_results}),
we cannot compare those results for \pb{} and \xcverifier{}. We mark them as $\timeout$ in~\pref{tab:pb_xcverifier_consistency}.

\subsubsubsection{Summary for \textbf{RQ2}:} The results of \xcverifier{} are
consistent with the \pb{} approach (\pref{tab:pb_xcverifier_consistency}). For
DFA-condition pairs for which \pb{} finds counterexamples, \xcverifier{} also
finds counterexamples in similar regions. For pairs where \pb{} does not find
counterexamples, \xcverifier{} either verifies the input domain, or partially
times out and partially verifies the domain.

\section{Related Work}
\label{sec:RelatedWork}

\subsection{Analysis of Density Functional Theory Approximations}

To the best of our knowledge, ours is the first work that uses formal methods to
verify correctness in the context of density functional theory, an important
scientific computing application. Prior work has used a testing-based approach,
which is more scalable but does not provide formal guarantees. As discussed
earlier, Pederson and Burke~\cite{Pederson2023} use a grid-search to
evaluate whether the DFA satisfies the DFT exact conditions. 
Lehtola and
Marques~\cite{lehtola2022many} show that many 
recent DFAs are numerically ill-behaved 
by studying their accuracy in computing 
the total exchange-correlation energy.

\subsection{Correctness in High Performance Computing}

Correctness in scientific
computing is recognized as a major challenge in
HPC~\cite{DBLP:journals/corr/abs-2312-15640,DBLP:journals/corr/GopalakrishnanH17},
which needs formal methods that address the unique challenges in this
domain. Progress to date includes the verification of mathematical properties in
a conjugate gradient solver, a finite difference stencil, and a mesh quality
metric~\cite{DBLP:conf/sas/HuckelheimLNSH18}, PDE
solvers~\cite{DBLP:journals/toms/BientinesiGMQG05}, specific properties of
CG~\cite{DBLP:conf/sbmf/MarcilonJ13} and LU
decomposition~\cite{lu-decomposition}, and the floating-point equivalence of
manually and automatically differentiated
code~\cite{DBLP:conf/sc/SchordanHLM17}. The above are orthogonal to our goal of
verifying exact conditions of density functional theory approximations.
Extending our approach of using \dreal{} to verify properties of 
other scientific computations is an interesting future direction.

\subsection{Analysis of Floating-Point Programs}

Many testing and analysis techniques for floating-point programs have been
developed in the past decade. The first set of techniques are general approaches
that aim to achieve high-coverage of numerical code
~\cite{DBLP:conf/pldi/FuS17,fpgen}, conduct differential testing of numerical
libraries~\cite{fpdiff}, and perform mutation testing of floating-point
expressions given a real specification~\cite{DBLP:conf/issta/JeangoudouxDL21}.
However, it has been shown that simply achieving high code coverage in numerical
programs does not uncover numerical issues in most cases~\cite{fpgen}.
Additionally, differential testing is not feasible for DFA implementations
because most of them are unique.

A large body of work focuses on performing automated error analysis of
floating-point
programs~\cite{DBLP:conf/oopsla/FuBS15,DBLP:journals/toplas/SolovyevBBJRG19,DBLP:conf/fm/SolovyevJRG15,DBLP:conf/sc/DasBGKP20,DBLP:journals/pacmpl/ZouZXFZS20,
DBLP:conf/cluster/DasTGK21,DBLP:conf/ipps/SinghKMPLV23,DBLP:conf/sas/AbbasiD23,DBLP:conf/ifm/LoharPD19,DBLP:conf/tacas/DarulovaINRBB18,DBLP:conf/fmcad/IzychevaD17,DBLP:conf/nfm/DamoucheM17}.
While some of these approaches provide sound error bounds of floating-point
programs, they suffer from important limitations with respect to
program size and control structures supported. In the absence
of techniques that can reason about floating-point error in non-trivial programs, a rich area of research in software testing has focused on how to
efficiently generate inputs that \emph{maximize} error in the output of a
program~\cite{DBLP:conf/ppopp/ChiangGRS14,fpgen,DBLP:conf/icse/ZouWXZSM15},
which can shed light on the potentially worst error a floating-point program
could incur. Similarly, work has proposed techniques to generate inputs that
trigger floating-point
exceptions~\cite{DBLP:conf/popl/BarrVLS13,DBLP:conf/sc/LagunaG22}. Calculating
error bounds of DFAs is orthogonal to our goal of verifying physical
and numerical properties of their implementations.

Other existing work on floating-point programs 
has explored finding function input ranges, also referred to as
regimes, with the purpose of improving the accuracy of floating-point
expressions~\cite{DBLP:conf/pldi/PanchekhaSWT15} or optimizing floating-point
efficiency~\cite{DBLP:journals/tecs/RabeID21}. These approaches are based on
either estimating error based on dynamic input sampling, or statically
performing an error analysis, which have their own limitations, as
described earlier.

\section{Discussion}
\label{sec:Discussion}

\subsection{Improving Scalability of the Solver} 

In our experiments, \xcverifier{} was unable to verify any of the exact conditions of the SCAN
functional, which has been designed to satisfy all known exact conditions. SCAN
is significantly more complex than the other functionals we considered, and also
involves the use of transcendental functions such as $\exp$ and $\log$. 
This
causes \dreal{} to time out even for the relatively simple $\ec$ non-positivity condition, 
and even when the input domain is reduced $32\times$.
It would be
interesting to investigate approaches to improve the performance of the solver
so that it can tackle the SCAN functional. Apart from its popularity, the SCAN
functional will serve as a fascinating use case: there is a progression of
DFAs---rSCAN, r++SCAN, r\textsuperscript{2}SCAN,
r\textsuperscript{4}SCAN---proposed in the literature that were designed with
different adherence to exact conditions to improve the numerical stability of
the original SCAN
functional~\cite{bartok2019regularized,furness2020accurate,furness2022construction}.

\subsection{Expanding to More DFA-Condition Pairs}

Our evaluation demonstrated the robustness of the \xcencoder{} to handle a wide
variety of DFAs and exact conditions. The ultimate goal of our research is to be
able to analyze all the 500+ functionals in \libxc{} for all known DFT exact
conditions. Future work will continue to expand our evaluation, and will aim to
integrate our verification tool into \libxc{}, e.g., as part of the continuous
integration (CI) for \libxc.

\subsection{Numerical Issues With DFAs}
\label{sec:DFTPreliminaries-Numerical}

Apart from verifying known exact conditions for DFA implementations, it would be
interesting to analyze numerical issues of the implementations with the goal of
using formal methods to find and fix numerical issues in DFA implementations.
This is a challenging problem involving reasoning about floating points and
dealing with transcendental functions like $\sin$, $\log$ and $\exp$. The
functional forms of DFAs themselves can also be a source of numerical issues.
Some DFAs include different functions that apply to different input domains, and
must ensure continuity when switching from one domain to another. Additionally,
the parametrization of the DFA may cause issues. Even in the simple case of the
Local Density Approximation (LDA), the Perdew-Zunger~\cite{perdew1981self}
parametrization of the results of Ceperley and Alder~\cite{ceperley1980ground}
includes potentially inaccurate numerical constants that lead to discontinuities
of the exchange-correlation energy at a given matching point.

The functional form of a DFA may also make it sensitive to inaccuracies in its
input data. While a given implementation of a DFA may yield correct answers for
an exactly known (e.g., exponential) density, it may result in large numerical
errors if the input density is noisy, or if the density and its gradient are not
numerically consistent. This is particularly problematic in regions of low
density, e.g., a point far from a molecule placed in vacuum. Such large errors may lead
to inaccurate energies or slow convergence in the solution of the Kohn-Sham
equations.
For example, the sensitivity of the SCAN
functional requires the use of extremely fine grids to represent the electron
density in order to avoid large numerical errors. This led some authors to
modify the SCAN functional to avoid this numerical issue, resulting in a
slightly different DFA~\cite{furness2020accurate}.
In other cases, an analytical reformulation of a DFA is used to avoid numerical
issues~\cite{wu2012simplified} without modifying it. However, these fixes are
ad~hoc, and there is no known general recipe for avoiding the numerical issues of a
DFA.

\section{Conclusion}
\label{sec:Conclusion}

This paper presented \xcverifier{}, an approach for verifying whether a DFA
implementation satisfies the DFT exact conditions. \xcverifier{} automatically
encodes the DFA implementation from \libxc{} and a given exact condition into a
\dreal{} formula, symbolically performing any required derivative
calculations. It uses the \dreal{} solver to verify whether the condition is
true or find a violation to it for a given input domain. \xcverifier{} also implements a
domain-splitting technique to improve performance, reducing solver timeouts,
and isolating the input regions where the condition is satisfied or
violated. 

We evaluated \xcverifier{} by verifying seven exact conditions (from Pederson
and Burke~\cite{Pederson2023}) for five popular DFAs. \xcverifier{} was successfully able to verify
or find a counterexample for 13 out of the 31 (valid) DFA-condition pairs, and
it was able to partially verify 
an
additional seven pairs. 
However, it timed out for 11 pairs, which included all the conditions for the SCAN functional. 
We found that the results of the \pb{} approach, which used grid search 
to check DFT exact conditions, were largely consistent with those of \xcverifier{}.
These results demonstrate promise and future challenges of using 
formal methods for DFT.

\bibliographystyle{IEEEtran}
\bibliography{main,dft,rubio,thakur,francois,fm}

\end{document}